\documentclass[a4paper,fleqn]{cas-sc}

\usepackage[authoryear]{natbib}
\usepackage{multirow}
\usepackage{pgfplots}
\usepackage{pgfplotstable}
\usepackage{subfigure}
\usepackage{tikz}

\def\tsc#1{\csdef{#1}{\textsc{\lowercase{#1}}\xspace}}
\tsc{WGM}
\tsc{QE}
\tsc{EP}
\tsc{PMS}
\tsc{BEC}
\tsc{DE}

\begin{document}
\let\WriteBookmarks\relax
\def\floatpagepagefraction{1}
\def\textpagefraction{.001}
\shorttitle{Learning driving style embedding for driver identification}
\shortauthors{Lin Lu et~al.}

\title [mode = title]{Learning driving style embedding from GPS-derived moving patterns for driver identification}

\author[1,2]{Lin Lu}[orcid=0000-0002-2803-6297]
\cormark[1]
\ead{linklu@whut.edu.cn}
\credit{Methodology, Writing - Original draft preparation, Proofreading}



\address[1]{College of Computer and Information Technology, China Three Gorges University, Yichang, 443517, China}

\begin{abstract}
Learning fingerprint-like driving style representations is crucial to accurately identify who is behind the wheel in open driving situations. This study explores the learning of driving styles with GPS signals that are currently available in connected vehicles for short-term driver identification. First, an input driving trajectory is windowed into subtrajectories with fixed time lengths. Then, each subtrajectory is further divided into overlapping dynamic segments. For each segment, the local features are obtained by combining statistical and state transitional patterns. Finally, the driving style embedded in each subtrajectory is learned with the proposed regularized recurrent neural network (RNN) for short-term driver identification. We evaluate the impacts of key factors and the effectiveness of the proposed approach on the identification performance of 5 and 10 drivers. The results show that our proposed neural network structure, which complements movement statistics (MS) with state transitions (ST), provides better prediction performance than existing deep learning methods.
\end{abstract}


\maketitle

\section{Introduction}
Automobile sensor data have promoted the popularity of data-driven methods to better understand driving behaviors. Vehicle driving data indicate drivers’ unique driving habits or styles~\citep{AutomobileDriverFingerprinting, 8863987, halim2016profiling}, exactly as their fingerprints indicate their unique identities. This type of digital fingerprint plays a key role in driver identification tasks in many intelligent transportation applications~\citep{9430766}, such as risky driving recognition~\cite{WANG2022106589}, theft protection~\citep{8863987}, personalized driver assistance systems~\citep{Lin2014} and even autonomous driving~\citep{7139555}.

The primary objective of the driver identification task is to determine the identity of the person behind the wheel~\citep{NASRAZADANI2022105}. With an increasing focus on privacy protection~\citep{AutomobileDriverFingerprinting} and application cost savings~\citep{https://doi.org/10.1049/iet-its.2014.0248}, using sensor data, i.e., data collected through CAN-bus, GPS, and smartphones to understand and recognize driving styles, which uniquely identify drivers, has drawn considerable attention. GPS-based driver identification methods are especially popular since GPS sensors have lower costs and are widely equipped in vehicles and mobile phones. Therefore, this research focuses on developing a characterization of driving style with GPS signals that can be used to model and understand human driver behavior.

However, the raw GPS traces themselves cannot directly distinguish between drivers with different driving styles. Hence, the primary challenge is how to transform drivers’ variable-length GPS trajectories into an appropriate structure that can characterize driving activities and corresponding spatiotemporal dynamics. Recently, deep learning has been introduced as the dominant way to map driving behavior to a user-specific embedding, from which many downstream tasks can also take advantage. However, none of the related proposals has adequately addressed this challenge because of the lack of generalizability in their structures. For example, \cite{Dong2016} proposed learning from a low-level statistical feature matrix using a convolutional neural network (CNN) and long short-term memory (LSTM), but movement statistics (MS) in the matrix ignored temporal dependencies that reflect vehicle state transition (ST) relationships. In contrast, the ST view was addressed by~\cite{Wang2018}, who proposed learning a latent driver representation from two points of view: transition duration and frequency. These two views depend on predefined driving states; however, they ignore the intensity that reflects the magnitude of neighboring state changes. As recognized, moderate and intense braking is treated as the same braking event without differentiation. In other words, the moving/driving patterns involved in the driver identification problem have not yet been fully considered.

We also note that the previous literature has primarily focused on the passenger car (sedan), while the driving behaviors of large truck drivers were not specifically addressed. The functions of trucks compared to those of sedans result in unique driving patterns that require more comprehensive study. For truck driving behavior, the state changes responding to a driver's operations may not be as sensitive as those of a sedan due to their differing load and power capabilities. This fact poses a greater challenge to effectively learning compressed and generalized representations to differentiate between the styles of various drivers. Therefore, this paper is particularly interested in the driver identification problem involving different types of vehicles, in which only GPS traces collected in real time are provided.

To this end, it is necessary to study the different driving patterns of sedan and truck drivers and develop a more generalized identification method. Therefore, this study first explores the integration of statistics describing a kinematic segment with the state views to represent the associated driver's skill. Given a driving trip, we map the GPS points into fixed-length segments containing the MS and ST patterns. Then, a novel neural network framework is presented to automatically learn the associated driving style representation and further predict the identity of the driver for a subtrajectory containing several segments. We demonstrate how our approach helps improve the accuracy in the driver identification problem. In summary, our work makes the following contributions.

In this paper, we address the driver identification task to alleviate the challenges outlined above. The main contributions of the present work can be summarized as follows:
\begin{enumerate}
\item[-] A robust moving pattern is proposed by incorporating statistical and state transition views to enrich the driving behavior information in the driving segment of GPS traces.

\item[-] A residual recurrent neural network-based identification framework is designed to learn a high-level representation of the driving style from low-level sequential patterns and perform driver identification.

\item[-] To demonstrate the robustness of the locally fused driving pattern and the performance of the identification models, experiments are carried out on GPS data from both sedans and trucks. The results show that the driving patterns are different between truck and passenger car drvers, and our solution is more robust and outperforms existing driver identification methods.
\end{enumerate}

The remainder of this manuscript is structured as follows. In Section~\ref{sec_2}, studies related to the problem of driver identification are briefly introduced, and the literature gap is discussed. 
Section
~\ref{sec_3} first formulates the problem of concern and then introduces our solution in detail. In Section~\ref{sec_5}, our experimental process is described, and the results of various experiments are shown and compared. Section~\ref{sec_6} concludes the main contribution of our work and presents a future research focus.

\section{Related work}
\label{sec_2}
Driver identification can be formulated as a classification problem in machine learning (ML). To solve this problem, considerable work has been done with the aim of characterizing driving behavior and then identifying the correct driver.

Early work resorted to conventional ML algorithms to recognize drivers. The work of~\cite{1520171,Miyajima2007,ozturk2012driver} revealed that a driver's operation signals can be combined with Gaussian mixture models (GMMs) to help match cars and drivers. The main difference in their work was the inclusion of more signals and manually selected features. One shortcoming of these methods is that they use driving simulators as their primary data source, which offers rich signals but is limited in demonstrating and capturing the exact real-world driving experience and conditions. With the development of the Internet of Vehicles, real-world driving data can be collected, attracting more prospective methods. Therefore, \cite{9724165} validated the GMM by identifying drivers who naturally drove on real-world roads. Extreme learning machines\citep{7313563} and support vector machines (SVMs)~\citep{qian2010support} were also adopted to identify drivers. Later, \cite{li2019driver} considered K-nearest neighbors (KNN), random forest (RF), AdaBoost, and multilayer perceptron (MLP) models in their study. The results showed that the RF algorithm offered the best performance. When focusing on real-time identification, \cite{Jafarnejad2018} used statistical and cepstral features to develop a window classification function with a set of traditional ML methods. They proposed two decision functions to obtain a final prediction at the trip level. The experiments yielded favorable identification rates. \cite{9108547} proposed a driver identification method based on a wavelet transform by performing a driving pattern analysis for each driver. They compared the performances of three different ML algorithms, namely, SVM, RF, and extreme gradient boosting, to perform driver identification. Despite the good classification performance, the above methods for modeling driving styles were based mainly on hand-crafted features~\citep{Lin2014,Dong2016}. It is challenging to transform variable-length driving traces into a fixed-length vector that characterizes driving behavior and manually defines driving style using traditional feature engineering. In particular, several drawbacks have been encountered, including the following: 1) Considerable efforts are needed to select features before feeding into ML algorithms; 2) the temporal relation within a sequence of driving operations is not considered; and 3) the best descriptors of driving patterns may change given different data or vehicles, e.g., sedan drivers may have different driving patterns from those of truck drivers.

To overcome the drawbacks of hand-crafted features needed by traditional ML algorithms, deep learning-based methods have recently been proposed to automatically learn representative features (or embeddings). Using a time series of controller area network (CAN) bus signals, \cite{Jeong2018} used techniques such as a one-dimensional (1D) convolutional neural network (CNN), normalization, special section extraction, and postprocessing to improve the accuracy of the identification. 1D CNNs were also adopted by~\citep{hu2020driver}. On the other hand, RNNs were considered to address the second drawback by implementing LSTMs~\citep{girma2019driver}. Furthermore, \cite{moosavi2021driving} proposed a model that combined CNN and LSTM for driver identification (D-RCNN), and the experimental results showed that the performance of the D-CRNN is better than that of the CNN and its regularized version of the autoencoder (ARCNN).
\cite{9145829} proposed a driver embedding learning architecture that included a 50-layer residual network (ResNet-50) followed by two stacked GRUs based on smartphone accelerometer signals. The architecture was applied to driver identification and driver verification problems. Siamese-LSTM was recently considered by~\citep{8813795,9716898} to learn driving style embeddings by treating driver identification as a matching problem. The experimental results of \cite{9716898} showed that their methods achieve state-of-the-art performance over simple GRUs, LSTMs, feedforward NNs, etc.
Other research, such as \cite{LU2022117299}, which focused on the problem of identifying drivers with a few shots, is beyond the scope of this study.

Focusing solely on GPS data, \cite{Dong2016} proposed several networks that could learn driving style characteristics. Using driver identification as a testing task, they developed several deep neural network architectures, including 1D CNNs and RNNs, and studied the performance of these architectures in terms of learning suitable driving-style representations from transformed data input. An experiment on empirical taxi data showed that RNNs outperformed CNN and gradient-boosting decision tree (GBDT) methods. The authors of the aforementioned study subsequently proposed an autoencoder-regularized RNN to generalize their prediction algorithm and scale it for use with unseen driver data~\citep{Dong2017}. In addition, \cite{chowdhury2018investigations} presented an approach for driver identification using GPS data solely from smartphones. As we observe in these studies, the descriptive statistics of the GPS signals are used to characterize the driving patterns and then fed into the proposed networks. Examples of MSs include the average speed, the maximum acceleration, and the standard deviation of the jerk. Using this pattern transformation, GPS traces of various lengths can be transformed into a fixed-length structure that can be accepted by neural network models.

Another kind of driving pattern being considered is driving events or states. A driving event is generally understood to be a maneuver that occurs during the driving task, such as acceleration, deceleration, turning and lane change \citep{PENG2020107276}, that can be used to identify the corresponding driving style. A temporal combination of sensor signals defines driving events/states. In the short term, a vehicle is limited to acceleration, braking, and turning events. Along a real, uncontrolled road, a short-term event usually lasts only a few seconds. The duration of such an event can also be used to measure aggression because harsh driving events often last for shorter periods than normal driving events. In contrast, a long-term event is a combination of short-term events that has a higher level of semantic complexity. Complex maneuvers can be a combination of acceleration, braking, and turning events. For instance, a U-turn can be classified into two turning events. Using driving states derived from raw GPS data, \cite{Wang2019} constructed a sequence of ST graphs to characterize driving behaviors. They then proposed a GRU-based network to learn spatial and temporal embedding and applied their proposed method to predict driving scores and detect risky areas for driving. \cite{Chen2016} developed a stacked denoising autoencoder architecture to extract features from data on human mobility and linked the deep representation obtained to the level of risk posed by traffic. These deep learning models can efficiently reflect a driver's fine-grained driving habits and can improve the accuracy of driver identification.

However, the above deep learning methods using GPS data considered only one type of driving pattern. To better characterize the driving behavior of both sedan and truck drivers, comprehensive driving patterns should be considered, and accordingly, an appropriate neural network structure is needed.

\section{Proposed solution}
\label{sec_3}
\begin{figure}[htbp]
  \centering
  \includegraphics*[width=0.99\textwidth]{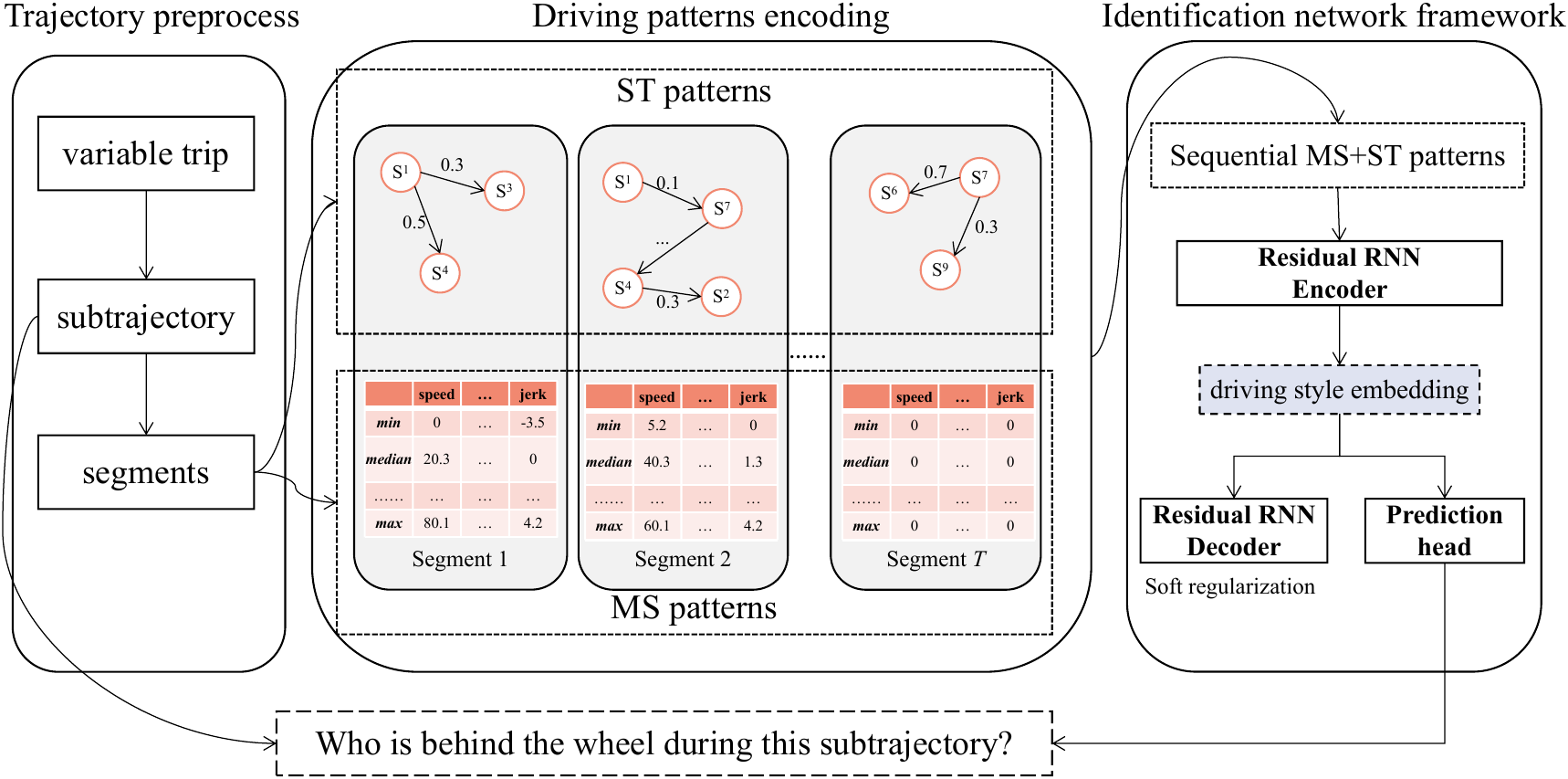}\\
  \caption{Proposed framework for short-term driver identification}\label{FIG:framework}
\end{figure}

First, we define some symbols in Table~\ref{tab:symbols}. Next, we formulate the identification of drivers in real time or in the short term as the task of predicting which driver is behind the wheel during any subtrajectory. We limit the target drivers to a finite set and then consider driver identification as a classification problem.
The goal of driver identification is to associate a GPS trace $\mathbf{x}$ with the corresponding driver label $y$. The learning process depends on the learning strategies used. To solve this problem, we propose a novel solution, as illustrated in Fig.~\ref{FIG:framework}, which consists of three stages.

\begin{table}[!htbp]
    \centering
    \caption{The symbols defined in this study}
    \begin{tabular}{clcl} \hline
        Symbol & Description & Symbol & Description\\ \hline
        $Tr$ & GPS trjactory & $s^i$ & the $i^{th}$ driving state\\  
        $p$ & GPS point & $I(\cdot,\cdot)$ & transition intensity function\\  
        $T$ & timesteps & $(\boldsymbol{x}, y)$ & single traninig example\\ 
        $M$ & feature dimension & $c$ & number of classes \\
        $v$ & speed & $\boldsymbol{q}$ & prediction output vector \\
        $a$ & acceleration & $\tilde{y}$ & predicted label \\
        $j$ & jerk & $n$ & number of training examples\\
        $b$ & bearing & $L_c$ & classification loss\\
        $ba$ & angular speed & $L_c$ & reconstruction loss\\
        $bj$ & angular jerk & $L_u$ & regularized loss \\
        $L_f$,$L_s$ & sliding time window length & $\lambda$  & regularization weight \\  \hline
    \end{tabular}
    \label{tab:symbols}
\end{table}

In the first stage, any variable-length trip of a driver is divided into several short-term subtrajectories. We focus on predicting the correct driver ID given a subtrajectory. Then, any subtrajectory is further segmented by a sliding window into $T$ segments. Each segment is closer to real time, but the data collected during the segment may not support a reasonable prediction, especially when the segment contains an idling stop or a constant driving state. Section~\ref{sec:preprocess} details this stage.

These segments are used to produce a sequence of driving patterns in the next stage. The stage of driving pattern encoding defines two patterns for any segment, namely, the state transition (ST) pattern displayed in a graph structure and the movement statistic (MS) pattern in matrix view. The two types of patterns are concatenated into one local feature vector with dimensions $M$. By doing so, the GPS points of a subtrajectory are transformed into sequential MS + ST patterns (denoted $\mathbf{x}$) with a shape $(T, M)$, where $T$ and $M$ denote time steps and dimensions, respectively. This stage is described in detail in Section~\ref{sec_pattern}.

The aim of our solution is to learn unique driving styles to improve driver identification performance. Therefore, in the last stage, a general identification network framework is introduced to effectively handle sequential inputs. A residual RNN encoder first maps the sequential patterns to an embedding vector as the driving-style representation for a subtrajectory. Then, a residual RNN decoder is used for soft regularization in the training process. Finally, the driving-style embedding is fed to the prediction head to output the driver label. The typical choices for building the RNN architecture are LSTM~\citep{doi:10.1162/neco.1997.9.8.1735} and GRU~\citep{Chung2014}, which are both used in the experiment section.

\subsection{Trajectory preprocessing}
\label{sec:preprocess}
\begin{figure}[!htbp]
  \centering
  \includegraphics*[width=0.99\textwidth]{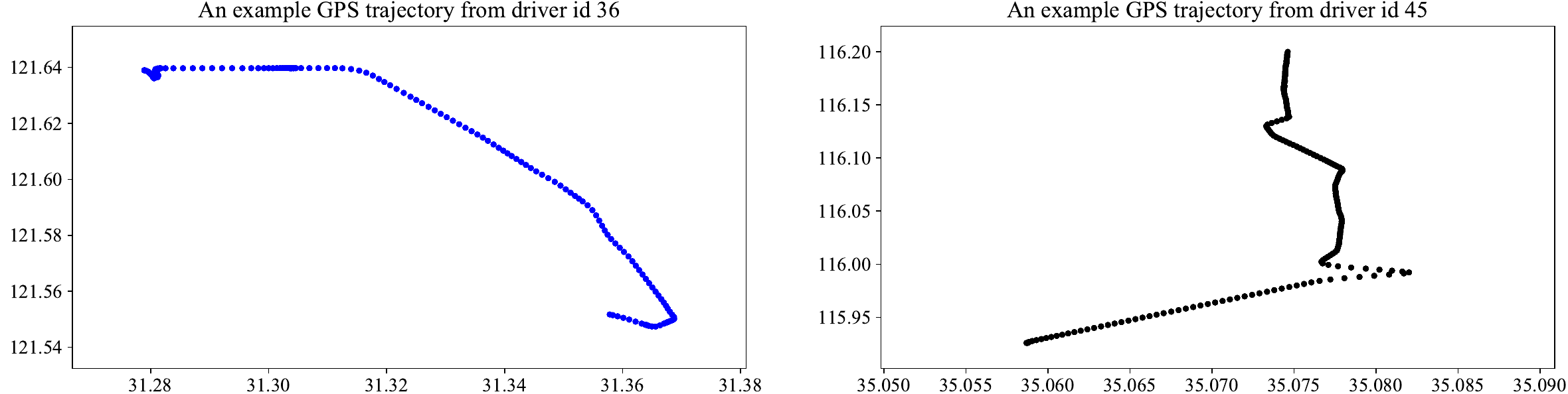}\\
  \caption{
GPS trajectories
of two example trips from drivers with IDs 36 and 45.}\label{traj}
\end{figure}

Given a running vehicle that belongs to a fleet that contains $N$ drivers, this vehicle incrementally produces a GPS trajectory denoted $Tr=<p_1, p_2, ...>$, where $p_i$ is a GPS point sampled at regular frequency. For illustration purposes, we randomly choose two trips assigned to drivers with IDs 36 and 45 from the truck fleet data set in Section~\ref{dataset}. The longitude and latitude of the two trajectories are shown in Fig.~\ref{traj}, which shows different driving behaviors that can be differentiated.

\begin{figure}[htbp]
  \centering
  \subfigure[Driver with ID 36] {
      \includegraphics*[width=0.45\textwidth]{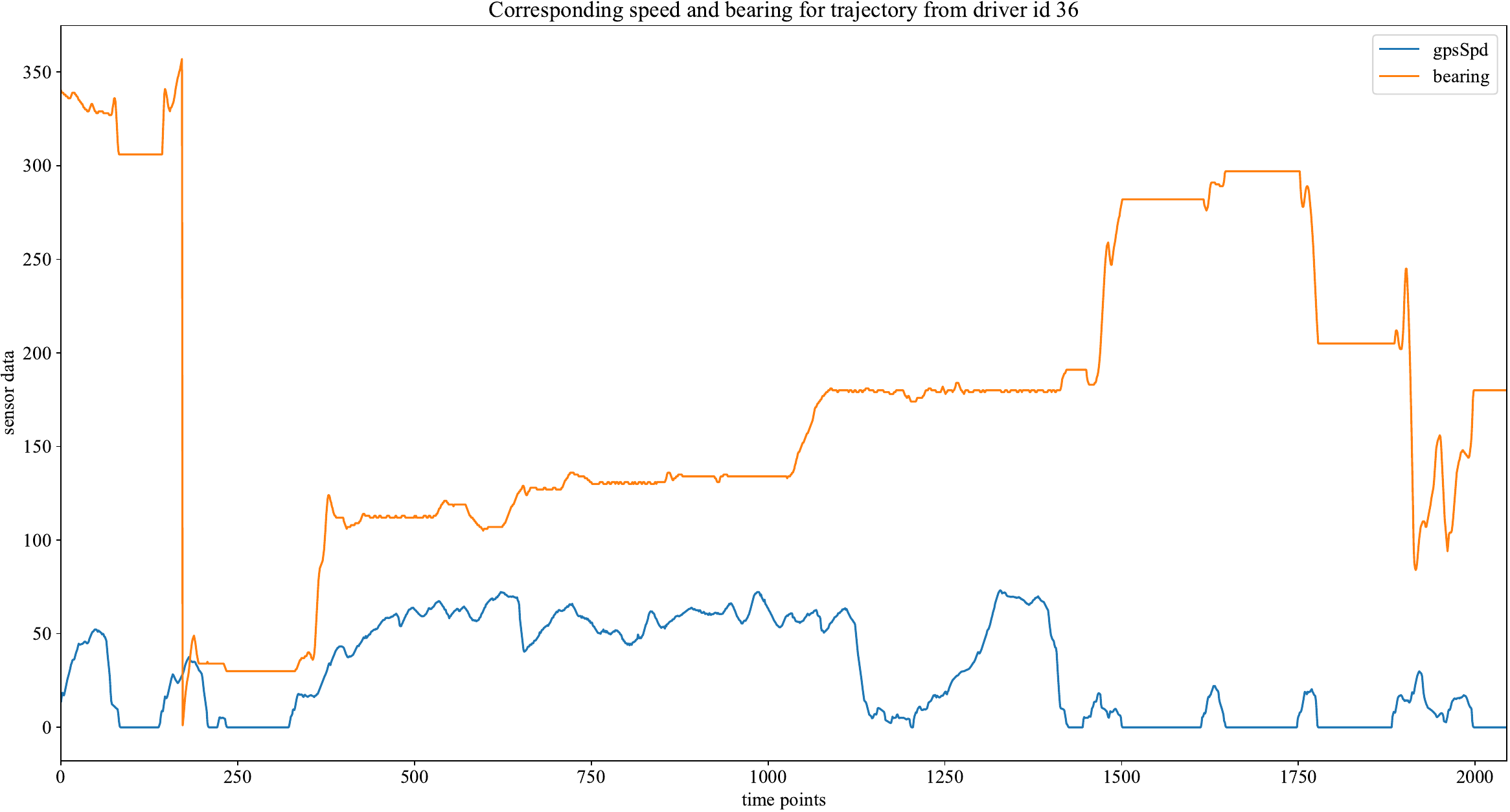}
  }
  \subfigure[Driver with ID 45] {
      \includegraphics*[width=0.45\textwidth]{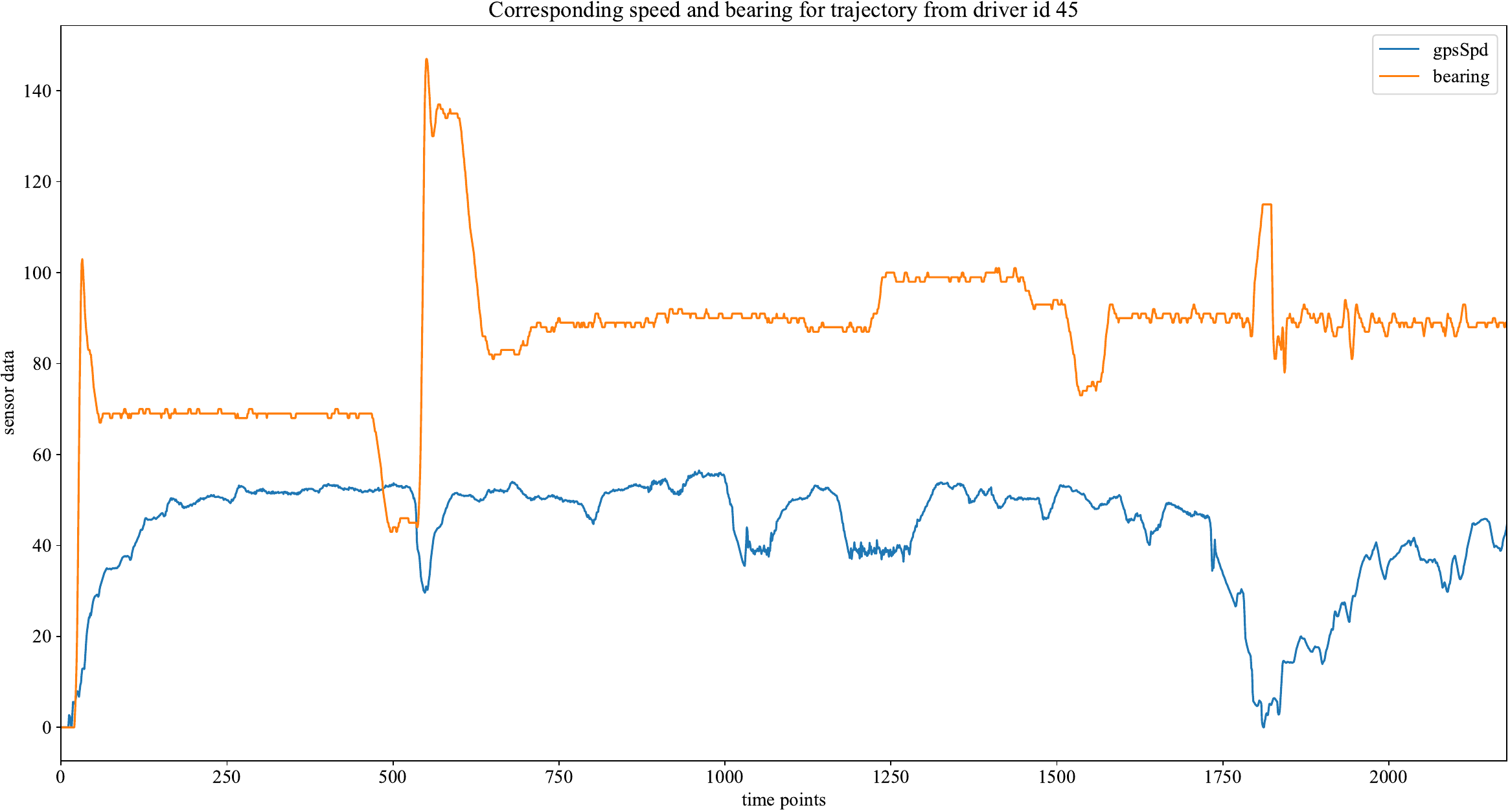}
  }
  \caption{Time series of the GPS-determined speeds and bearings corresponding to the two sampled trips.}\label{signal_series}
\end{figure}

However, due to the growing privacy concerns for position information exposure, we consider GPS-measured "speeds" and "bearings" as primary sources rather than utilizing GPS coordinates directly. According to Fig.~\ref{traj}, the line plots of the speeds and bearings derived from the GPS data are shown in Fig.~\ref{signal_series}. The two drivers operate on different routes and are therefore likely to exhibit different driving behaviors, as reflected by the GPS speed and bearing.

Moreover, the speed and bearing at a given point may not be sufficient to reflect certain driving behaviors, as they are usually conditioned on the previous two or three points. Therefore, we further define the concept of a \emph{trajectory point} to characterize the changing state of a vehicle at any time.

\textbf{Trajectory point}. For any GPS record at time $t$, we take the raw values and the first-order and second-order derivatives of the speed and bearing with respect to time. The first-order derivative of speed with respect to time is usually called acceleration, and the second-order derivative of speed is called jerk. The first-order derivative of the bearing with respect to time is called the angular speed, and the second-order derivative of the bearing is the angular jerk. Then, a trajectory point at time $t$ is defined as $p_t=[v_t, a_t, j_t, b_t, ba_t, bj_t]$, representing the corresponding speed, acceleration, jerk, bearing, angular speed, and angular jerk.

The definition of the trajectory point expands the speed and bearing to their first and second derivatives, providing more information about each point. To obtain the derivatives of $p_t$, three consecutive GPS records are taken. For the special cases of $p_0$ and $p_1$, zeros are utilized as masks.

\textbf{Fixed-length subtrajectory}. We first partition a given trajectory $Tr$ into smaller subtrajectories through a sliding \textit{ time window} with fixed length $L_s$. For short-term driver identification, we can build a model to predict the driver in one minute by setting $L_s=60$ seconds. For such a window of trajectory points, potential patterns can be discovered by observing the driver's behavior in different situations. For example, some drivers may go through a sharp corner quickly, while others may slow. Some drivers frequently accelerate, while others may never exhibit aggressive driving behavior. Such behaviors are interdependent in time, where the length of the period is $L_s$. We can assume that driving behavior at a given time depends on what has happened at the previous $L_s-1$ time points. Thus, we are more likely to discover driving patterns if we focus on the trajectory from a windowed perspective. We can also allow the machine to ``understand`` or ``define`` the driving style in this period.

\textbf{Kinematic segment}. To capture fine-grained movement characteristics, each subtrajectory is divided into several segments by an overlapping sliding window with a shorter length $L_f(L_f<L_s)$ and a shift of $L_f/2$. Each segment is then used for local feature extraction. Using the sliding-window method, the original variable-length trip is transformed into windows with a fixed number of segments. For example, given a driving trip that lasts 15 minutes (900 points), using $L_s = 256$ s and $L_f = 16$ s, the trip is divided first into $900/256=14$ subtrajectories. Each subtrajectory contains 256 points and is divided into $256/16*2=32$ segments. By treating each segment as a time step, we finally transform any given subtrajectory into 32 time steps.

\begin{figure}
  \centering
  \includegraphics*[width=0.96\textwidth]{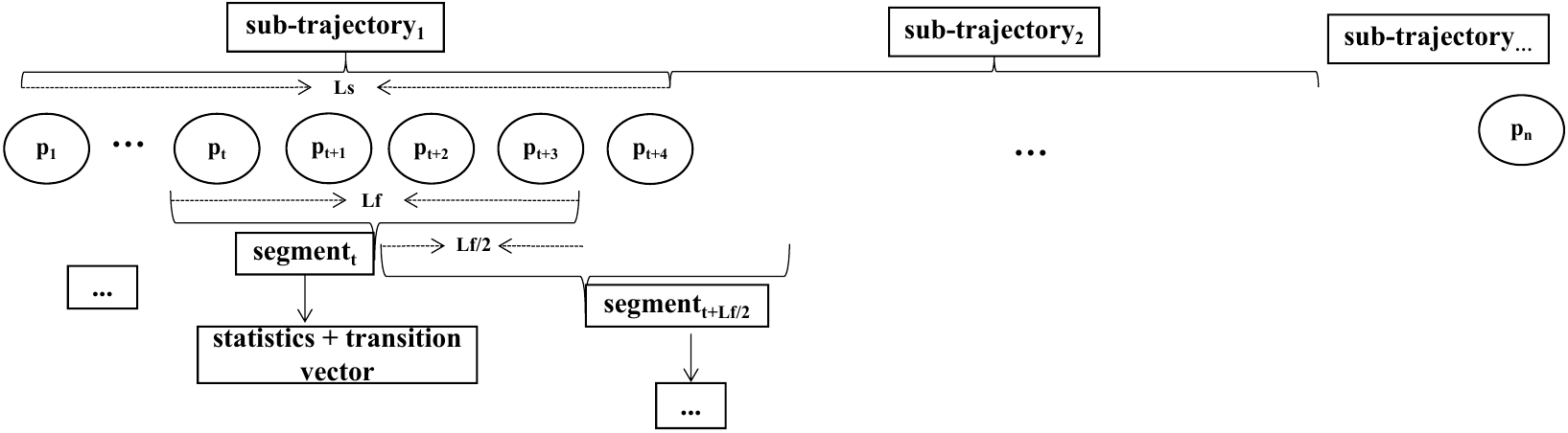}\\
  \caption{
Illustration of how the subtrajectories and
segments are obtained from the raw trajectory points for a trip. }\label{segment}
\end{figure}

As an illustration of this preprocess, Fig.~\ref{segment} shows how subtrajectories and segments are obtained given a time series of points for a trip. As we can see, the raw trajectory points are first divided into subtrajectories with fixed length $L_s$. Then, for each trajectory, a sliding window of length $L_f$ moves from left to right to produce dynamic segments. The step size to slide this window is set to half of the window length, that is, $L_f/2$. Next, we propose the transformation of the segment-level points into the domains of driving STs and MSs in Section~\ref{sec_pattern}.

\subsection{Driving pattern encoding}
\label{sec_pattern}
In this section, a novel pattern encoding strategy is introduced to transform a subtrajectory into the form of an input for the proposed neural network. Specifically, in Fig.~\ref{segment}, the proposed MS+ST pattern is extracted for each segment. In the following two sections, we introduce how the two patterns are obtained.

\subsubsection{State transitional pattern}
Limited-state machine theory is widely used to recognize the driving styles of human drivers and plan the motion of self-driving cars. Driving states convey explicit semantic meaning to represent drivers' skills. \cite{Chen2019} proposed a graphical method that first divides longitudinal acceleration into seven intervals corresponding to seven grades. For example, the code `3' represents acceleration values close to 0 ${m/s}^{2}$, that is, cruising. Grading increases with acceleration. As a result, longitudinal movements can be categorized into seven codes that represent different moving states. Previous studies \citep{Wang2019,Wang2018} have used driving events related to speed and direction to build ST graphs to capture temporally changing dependencies. The authors showed that the given state of a vehicle is dependent on previous operations and can affect future vehicle movements. Accordingly, we propose to transform neighboring trajectory points into the ST domain to characterize continuous operations of a driver.

For each segment, this section introduces a simple rule-based method to detect explicit driving states for any point. We first identify two categories of driving operations: (i) longitudinal operations that include `acceleration', `deceleration', and `constant speed'; and (ii) lateral operations that include `turning right', `turning left', and `moving straight'. If ${b}_{t}$ $\mathrm{>}$ ${b}_{t-1}$, then the operation is `turning right'; if ${b}_{t}$ $\mathrm{<}$ ${b}_{t-1}$, then the operation is `turning left'; otherwise, the operation is `moving straight'. Similarly, for speed-related operations, if ${v}_{t}$ $\mathrm{>}$ ${v}_{t-1}$, the operation is `acceleration'; if ${v}_{t}$ $\mathrm{<}$ ${v}_{t-1}$, then the operation is `deceleration'; if {\textbar} ${v}_{t}$ - v${}_{t-1}${\textbar} $\mathrm{<}$ $\Delta v$, then the speed is `constant`. $\Delta v$ is a bounded value that can be 1 km/h or another reasonable value.

As a result, the combination of 3 lateral/direction-related and 3 longitudinal/speed-related operations yields a total of nine driving states, which are listed in Table~\ref{driving_states}. In this manner, each point is assigned to any of the nine states. For example, a sequence of driving states for a vehicle can be [$\mathrm{<}$acceleration, moving straight$\mathrm{>}$, $\mathrm{<}$constant speed, moving straight$\mathrm{>}$, \dots, $\mathrm{<}$ deceleration, turning right $\mathrm{>}$].

\begin{table}[htbp]
\caption{List of nine predefined driving states.}\label{driving_states}
\begin{tabular}{cc} \hline 
Index & States \\ \hline
1 & Acceleration while turning right \\ 
2 & Acceleration while turning left \\ 
3 & Acceleration while moving straight \\ 
4 & Deceleration while turning right \\ 
5 & Deceleration while turning left \\ 
6 & Deceleration while moving straight \\ 
7 & Constant speed while turning right \\ 
8 & Constant speed while turning left \\ 
9 & Constant speed while moving straight \\ \hline
\end{tabular}
\end{table}

\begin{figure}[htbp]
  \centering
  \includegraphics*[width=0.5\textwidth]{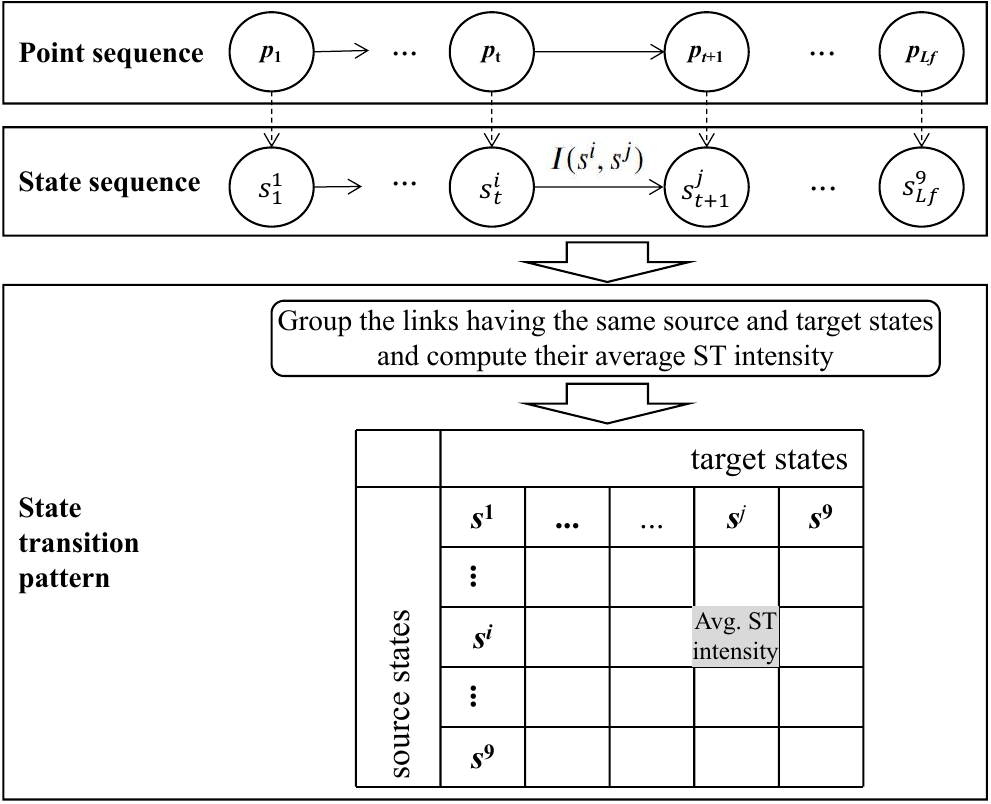}\\
  \caption{Illustration of the ST pattern extraction process.}\label{transition_repr}
\end{figure}

This sequence of driving states, transformed from the point sequence, is then converted into a driving ST graph (pattern). The extracted sequence of driving ST graphs is then used to characterize a driver's time-varying driving behavior. The process of producing such an ST pattern is shown in Fig.~\ref{transition_repr}. 

In Fig.~\ref{transition_repr}, $s^{i}_t$ represents the state of $p_t$ at time $t$, and $i$ is the index of the 9 predefined states in Table~\ref{driving_states}. For each pair of consecutive states, there is a direct link indicating the transition intensity. The intensity of the transition from source state $s^{i}$ to target state $s^{j}$ is defined by Eq.~\ref{intensity}.
\begin{equation}
\label{intensity}
\begin{array}{ccc}
Intensity & Speed \qquad \qquad Bearing & Frequency  \\
I(s^i,s^j)= & \sqrt{(s^i(v)-s^j(v))^2+(s^i(b)-s^j(b))^2} & +1
\end{array}
\end{equation}
Here, we let the transition intensity include the speed difference (namely, $(s^i(v)-s^j(v))^2$ and bearing difference $(s^i(b)-s^j(b))^2$), and the constant '1' is added if there is such a linkage in the state sequence.

\begin{figure}[htbp]
  \centering
  \includegraphics*[width=3.78in, height=2.61in, keepaspectratio=false]{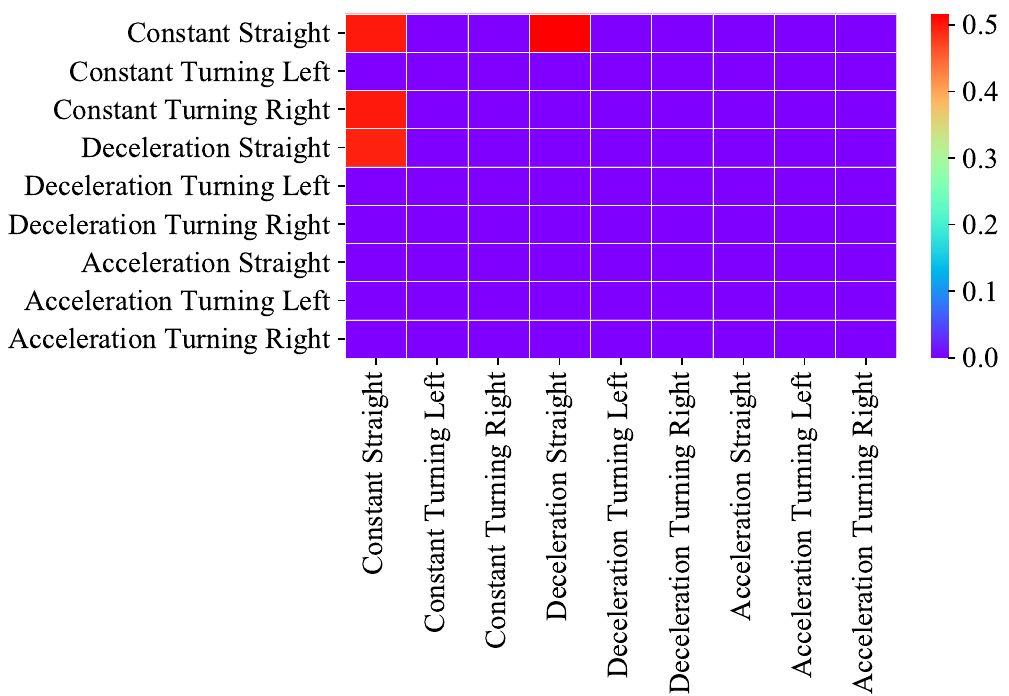}\\
\caption{Heatmap illustration of an ST pattern for a sample segment. For example, the first row indicates transitions from the `constant straight' state to itself and the `deceleration straight' state.}\label{heat_transition}
\end{figure}

After the ST values are obtained, an ST pattern in matrix form is constructed, where each entry denotes the average transition intensity. Specifically, for each pair $\mathrm{<}s^{i},s^{j} \mathrm{>}$, we sum the $I(s^{i},s^{j})$ values for all links that have the same source and target states in the state sequence; then, the cumulative sum of the transition intensities is divided by the occurrence of such transitions to obtain the average intensity. Finally, the ST matrices of all segments are normalized with MaxMinScaler, and therefore, the transition values are within the range 0 to 1. Fig.~\ref{heat_transition} shows the constructed ST matrix in a randomly selected segment, in which blue squares indicate no links between two states.

The transition intensities are notable for considering speed and direction changes, as well as the frequencies of such state pairs. Taking an acceleration event as an example, aggressive drivers exhibit frequent changes with large magnitudes at the position of the throttle pedal, whereas calm drivers show only small-magnitude changes with low frequency.

\subsubsection{Movement statistical pattern}
In this section, we consider the MSs of individual trajectory segments. According to Section~\ref{sec:preprocess}, each segment is a matrix with 6 rows and $L_s$ columns. Then, for each segment, a statistical feature matrix is constructed to capture the corresponding kinematic characteristics. The descriptive statistics of the movement measurements are calculated to characterize the dynamic pattern.

\begin{figure}[!htbp]
  \centering
  \includegraphics*[width=0.6\textwidth]{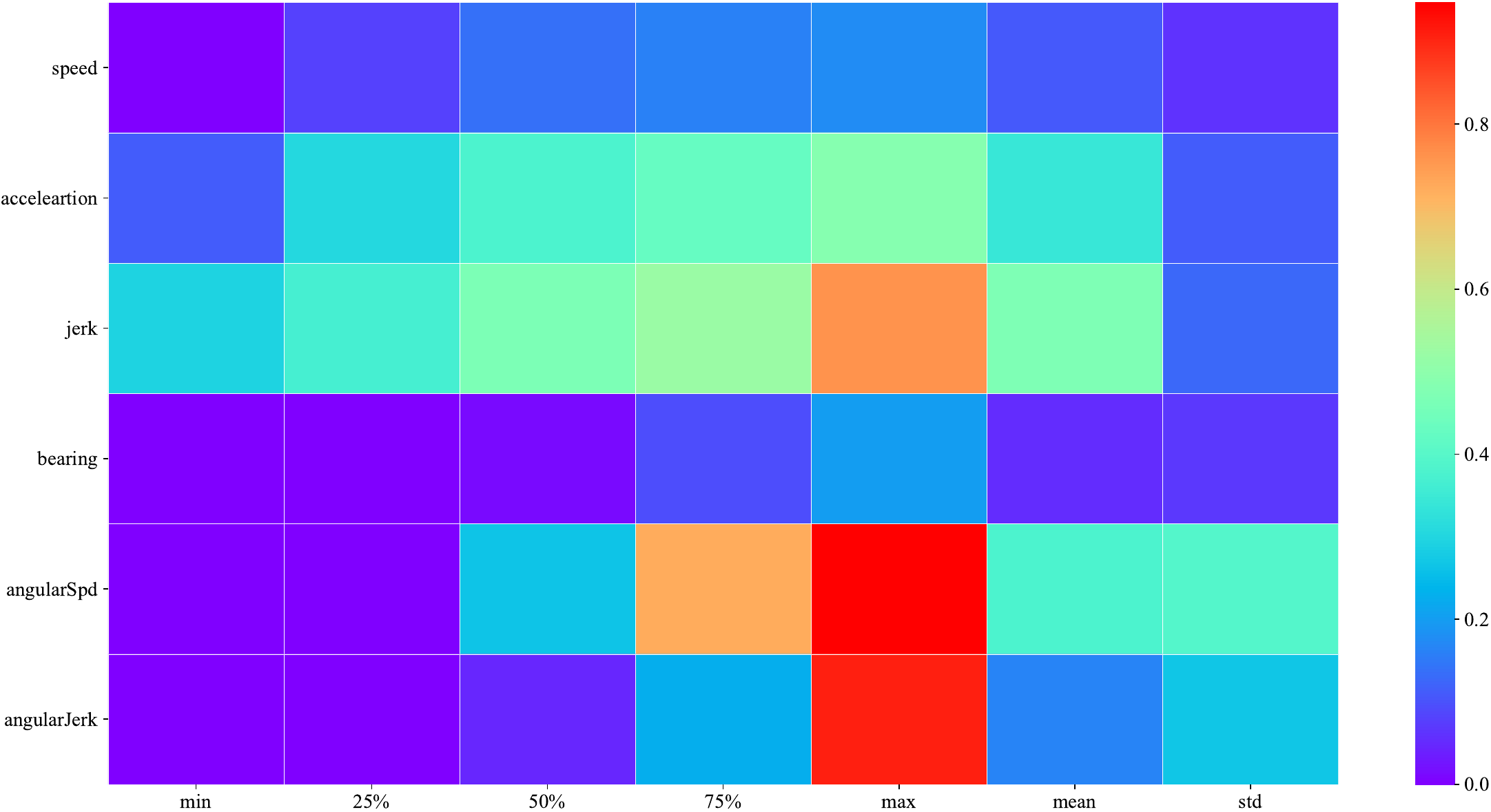}\\
\caption{The heatmap illustration of an MS pattern for a sample segment, where each row denotes the 7 statistics for one signal.}\label{st_matrix}
\end{figure}

Statistical features have been used extensively to analyze driving behavior. Previous studies have shown that the aggregation of short-term statistics along dynamic segments can help characterize the driving style of a driver on a trip \citep{Wang2019,Dong2016,Dong2017}. Consequently, we calculate the mean, minimum, maximum, 25\% quartile, 50\% quartile, 75\% quartile, and standard deviation to characterize the dynamics of each segment. Therefore, 7 statistics can be derived for each signal in $[v, a, j, b, ba, bj]$. An example of a segment-level statistical matrix is shown in Fig.~\ref{st_matrix}. Furthermore, $L_s$ is usually greater than 7, so the statistical matrix can be considered a method of reducing the dimensionality of each segment. Thus, for a given subtrajectory, a set of statistical feature matrices can be obtained. For example, using $L_s$ = 256 s and $L_f$ = 16 s, we define the matrices of size 42$\times$16 to characterize each subtrajectory.

\subsubsection{Driving pattern fusion}
In our consideration, the ST and MS patterns have their own drawbacks due to the partial information loss of a kinematic segment. Therefore, we propose merging both patterns to characterize the kinematic segments more thoroughly. The two matrices corresponding to both patterns together serve as the local features that will be fed to the proposed identification networks to learn driving style representations.

\textbf{Fused Pattern (MS+ST).} Pattern fusion represents the concatenation of the ST matrix and the MS matrix, but both matrices need to be flattened to vectors for fusion. Such an integration of the two patterns yields more information than either pattern alone can provide. For example, using the average speed, we can determine whether a constant driving state reflects a high speed. Statistics cannot show rapid movements, such as sudden braking, which can be reflected in the transition intensities of neighboring points.

\begin{figure}[!htbp]
  \centering
  \includegraphics*[width=0.6\textwidth]{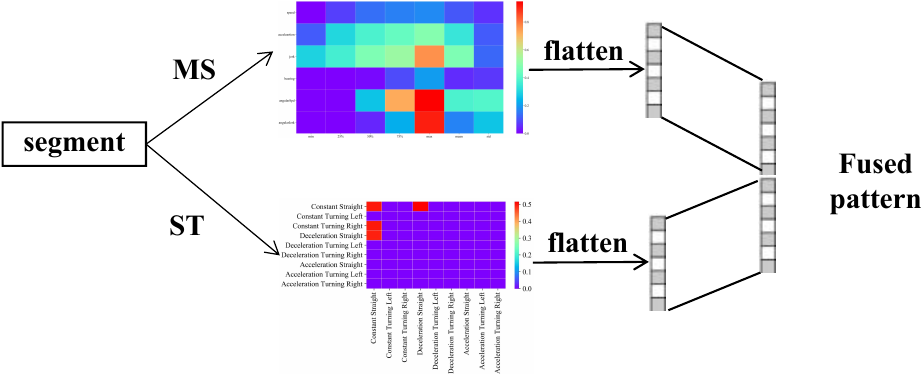}\\
\caption{An example of how to obtain the fused pattern for a driving segment.}\label{example_pattern_fusion}
\end{figure}

For a segment, Fig.~\ref{example_pattern_fusion} shows how to obtain the fused pattern vector given the MS and ST patterns displayed in Fig.\ref{st_matrix} and Fig.~\ref{heat_transition}, respectively. Finally, the sequential fused patterns and the correct driver labels at the subtrajectory level serve as training samples to train the driver identification model.

\subsection{Driver identification network}
\label{sec:networks}
\begin{figure}
\centering
\includegraphics*[width=0.4\textwidth]{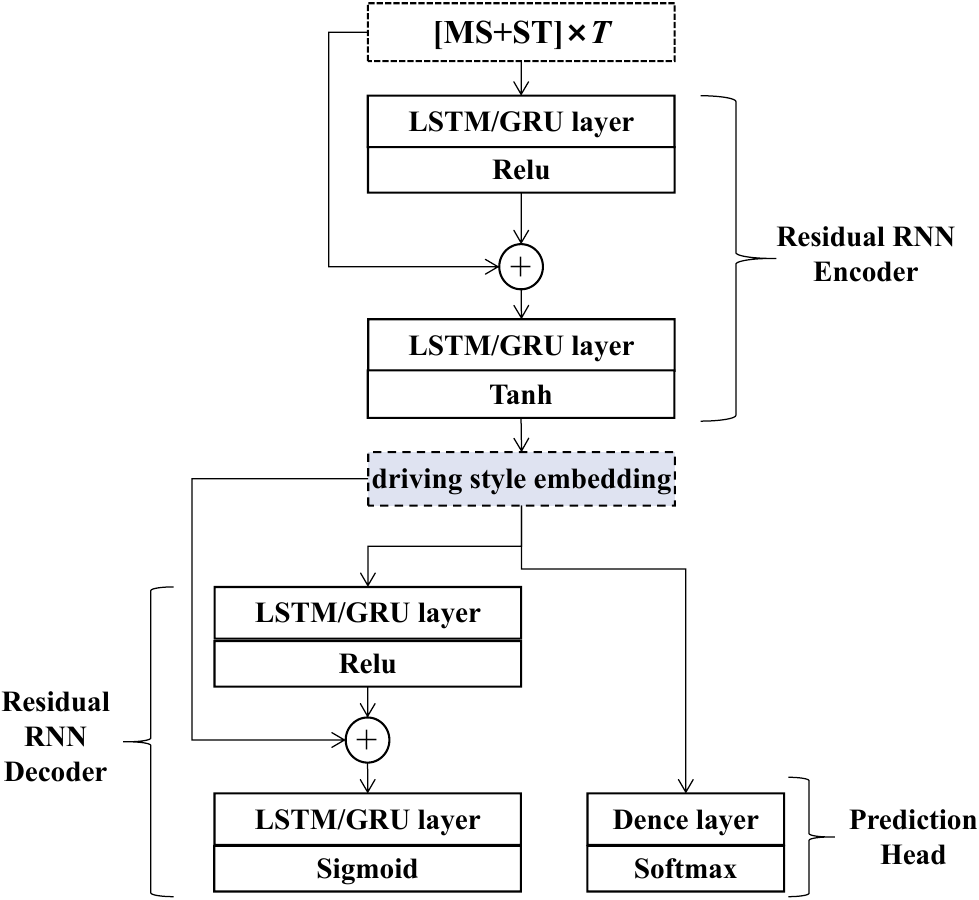}
\caption{The network structure of the proposed ResRNNARNet.}\label{united_net}
\end{figure}

Under the identification framework in Fig.~\ref{FIG:framework}, this section proposes a novel identification network structure. The structure is shown in Fig.~\ref{united_net} and consists of three components: a residual RNN encoder (also known as a feature extractor), a corresponding decoder for input reconstruction, and a prediction head. The RNN layers can be either GRU or LSTM layers. We introduce the three components one by one.

\subsubsection{Residual RNN encoder}
The residual RNN encoder aims to leverage the sequential properties of the input over time to better encode the dependencies between driving patterns. As shown in Fig.~\ref{united_net}, the encoder contains two recurrent blocks, and the residual/skip connection is used to add the input and output of the 1st block. Given a sequential MS+ST input with shape $(T,M)$, the first block maps the input to a rectified linear unit (ReLU)-activated sequence, whose shape is the same as the input. The second recurrent block then maps the additive output into a compressed state sequence. The last hidden state vector is treated as the driving style embedding, analogous to the digital fingerprint of a subtrajectory.

The adoption of residual connections is inspired by the residual LSTM proposed by \cite{kim2017residual}. Such a trick provides context from MS+ST patterns and further helps the next recurrent block identify sequential dependencies between the patterns and extract high-level driving style information. We found it beneficial in our application. Furthermore, residual connections help to provide better training for a deep network~\citep{he2016deep} and enable stacking of more recurrent blocks, although we only plot two here.

\subsubsection{Prediction head}
The prediction head includes a fully connected layer. It accepts driving-style embeddings as input, and softmax is applied to produce a distribution over the class labels. The number of classes (denoted \textit{ c}) is equal to the number of drivers in the training set. Given a training tuple $\mathrm{\{}\mathbf{x}$, y$\mathrm{\}}$, the predicted label $\tilde{y}$ is obtained by $\tilde{y}=\mathop{\mathrm{argmax}}_{j\in \{1,\dots ,c\}} \boldsymbol{q}_j$, where $\boldsymbol{q}$ is the probability vector produced by the softmax function in the prediction head.

The learning paradigm is typically a supervised learning network when only the RNN encoder and prediction head are used. This classic network is optimized by minimizing the loss between the predicted labels and the ground truth. The multiclass classification loss is defined as the standard cross-entropy loss, denoted $\mathcal{L}_c$.

\subsubsection{Residual RNN decoder}
As shown in Fig.~\ref{united_net}, the residual RNN decoder contains two symmetrical recurrent blocks to the encoder. Together, they form a residual RNN autoencoder. This autoencoder behaves in an unsupervised learning manner, mapping similar subtrajectories into nearby locations in the driving style representation space. This highlights the distinction between the proposed network and previously related supervised networks. Let the output decoded at time step \textit{t} be $\tilde{x}{}^{t}$. The autoencoder optimization process minimizes the loss between the reconstructed data and the original input. We use the mean square error metric to measure this loss, denoted $\mathcal{L}_r$.

The adoption of an additional decoder is mainly due to the following considerations:
1) From a generalization point of view, this decoder acts as a regularizer. Reconstruction of driving style embeddings is critical to learning better generalized driving style representations according to \cite{Dong2017, LU2022117299}. Therefore, the objective of using a specifically designed recurrent autoencoder structure is to regularize discriminative feature learning in a purely supervised network.
2) From an interpretable point of view, the decoder acts as a translator. It can reconstruct the abstract compressed driving style representation vector into the original readable driving pattern to show how it helps to identify specific drivers to make the system more easily accepted by the industry.

Finally, the overall objective function is defined as a partially weighted combination of the reconstruction and classification objectives. For this unified learning approach, the loss function is given by Eq.~\ref{loss_eq}:
\begin{equation}\label{loss_eq}
\mathcal{L}_u={\mathcal{L}}_c+{\lambda \mathcal{L}}_r
\end{equation}
where $\lambda \in (0,1)$ is a weight term used for soft regularization. The measurement methods for producing cross-entropy loss and reconstruction loss are different and are not on the same scale. If the reconstruction loss is large, the performance of the final classification accuracy will be greatly affected. Therefore, $\lambda$ is necessary but should be carefully tuned.

If the decoder is excluded, the learning of $\mathbf{x}$ is guided only by supervisory information about driver labels in the training data. However, given that the samples can be imbalanced among the drivers, the learning process without regularization may be prone to overfitting, exclusively learning the information in the training set. Hence, we must learn a representation that is more compact and delivers better generalization performance. As a straightforward extension, the decoding technique helps to obtain a better representation of driving style, which improves performance in terms of supervised learning. Minimizing ${\mathcal{L}}_u$ can help achieve this goal.

In summary, for our solution, we first obtain a sequence of locally fused patterns within a subtrajectory, and then the temporal relationship among these patterns is captured via our ResRNNARNet to produce better driving-style embeddings and identification accuracy. The advantages of our solution will be demonstrated in the next section.

\section{Experiments and discussion}
\label{sec_5}
In this section, we validate our proposed solution by solving the identification problems of 5 and 10 drivers on two data sets. To address the third drawback mentioned in Section~\ref{sec_2}, the two data sets are from passenger cars and logistic trucks.
For driving trajectory segmentation, $Ls$ is set to 60 seconds (1 minute).
Note that due to the varying trip lengths of the selected drivers, the preprocessed data are imbalanced. Drivers with long trips produce more training samples. Such imbalanced data are used to demonstrate the improved performance of our proposed unified model.

For our models and the comparative models, the dimension of driving-style embeddings is fixed to 100, the batch size is 256, the maximum number of iterations is set to 1500, and the other hyperparameters are properly tuned through extensive trials. In addition, we use both GRU and LSTM to instantiate the proposed ResRNNARNet. Hence, we have two models, denoted ResGRUARNet and ResLSTMARNet.
Then, 5-fold cross-validation is used, and the validation is repeated 5 times to produce the average accuracy metric. For each training fold, 15\% of the training data are retained for validation and early stopping.

\subsection{Data sets}
\label{dataset}
\href{http://rettore.com.br/prof/vehicular-trace/}{\textbf{Sedan Driving Data}}. \cite{rettore2018driver} from the Federal University of Minas Gerais carried out a case study to collect vehicle sensor data from two vehicles shared by fourteen anonymous drivers. Their driving data were collected at a frequency of 1 Hz. This public data set recorded the natural driving traces of volunteer drivers in the city and surroundings of Belo Horizonte, Brazil, in 2018. In our experiment, we only used data from vehicle 1 driven by ten drivers. Then, we randomly selected one trip for each driver, but only the GPS\_Speed\_Km and GPS\_Bearing signals were used to accommodate our setting. The preprocessed data are available via the link in the footnote~\footnote{https://gitee.com/cs\_icv/driver\_identification/tree/master/brazil\_fleet/data}.

\textbf{Truck Driving Data}. This is a private data set provided by a logistics fleet located in Shandong Province, China. There are 45 trucks of the same type driven by different hired drivers. The data acquisition frequency is 1 Hz. Each truck produces at least one trip belonging to a specific driver, and trips with durations less than 15 minutes are discarded. In particular, truck drivers drive freely in a completely uncontrolled and open driving environment; therefore, they can exhibit personalized driving styles. For privacy concerns, the corresponding raw GPS data are not publicly available, but the preprocessed data are available at the link in the footnote~\footnote{https://gitee.com/cs\_icv/driver\_identification/tree/master/truck\_fleet/data}.

Since not all vehicles operate simultaneously on the same day, it is not necessary to forecast the identities of all drivers in the short term. Therefore, we randomly select 5 and 10 drivers from the two data sets for the 5- and 10-driver identification problems. Taking the selected data samples for the 10-driver identification task as an example, we list their descriptive statistics in Table~\ref{tab:stats_sedan_data} and Table~\ref{tab:stats_truck_data}, where there are a total of 19534 and 16112 points, respectively. Although the two tables show similar average speeds, the acceleration and angular speed statistics show that the sedan driver group and the truck driver group have different behaviors in operating the acceleration pedal and steering wheel.

\begin{table}[!htbp]
\caption{The statistical description of sedan driving data}
\label{tab:stats_sedan_data}
\begin{tabular}{@{}lllllll@{}}
\toprule
      & speed  & acceleration & jerk   & Bearing & angularSpeed & angularJerk \\ \midrule
count & 19534  & 19534       & 19534  & 19534   & 19534        & 19534       \\
mean  & 33.36  & 0           & 0      & 155.44  & 2.62         & 3.25        \\
std   & 21.58  & 2.47        & 2.99   & 104.95  & 10.79        & 13.53       \\
min   & 0      & -51.34      & -52.99 & 0       & 0            & 0           \\
25\%  & 18.02  & -0.44       & -0.72  & 63.29   & 0            & 0.1         \\
50\%  & 36.28  & 0           & 0      & 160.7   & 0.4          & 0.6         \\
75\%  & 47.96  & 0.36        & 0.66   & 224.2   & 1.7          & 1.9         \\
max   & 116.18 & 50.76       & 58.5   & 359.92  & 179          & 179         \\ \bottomrule
\end{tabular}
\end{table}

\begin{table}[!htbp]
\caption{The statistical description of truck driving data}
\label{tab:stats_truck_data}
\begin{tabular}{@{}lllllll@{}}
\toprule
      & speed & accleration & jerk  & Bearing & angularSpeed & angularJerk \\ \midrule
count & 16112 & 16112       & 16112 & 16112   & 16112        & 16112       \\
mean  & 33.35 & -0.01       & 0     & 72.78   & 0.62         & 0.45        \\
std   & 22.94 & 1.02        & 0.94  & 81.63   & 1.92         & 1.05        \\
min   & 0     & -8.9        & -13.1 & 0       & 0            & 0           \\
25\%  & 8.7   & -0.4        & -0.3  & 21      & 0            & 0           \\
50\%  & 41.5  & 0           & 0     & 43      & 0            & 0           \\
75\%  & 54.1  & 0.3         & 0.3   & 94      & 1            & 1           \\
max   & 72.9  & 9.9         & 9.4   & 359     & 44           & 38          \\ \bottomrule
\end{tabular}
\end{table}

\subsection{Study on the influence of key factors}
\subsubsection{Segment length $L_f$}
The segment length determines the number of segments (or timesteps) that can be obtained from a subtrajectory. The segments are further transformed to a sequence of local patterns that are input into our models. We let $L_f$ be chosen in [10, 15, 20, 25, 30] and obtain the average classification accuracy of the two models on two identification tasks.

\begin{figure}[htbp]
\centering
\subfigure[ResGRUARNet]{
\begin{tikzpicture}[xscale=0.75, yscale=0.6]
  \begin{axis}[ybar,
  symbolic x coords={10,15,20,25,30},xtick=data,
  legend pos=outer north east]
  \addplot coordinates{(10,49.75) (15,51.25) (20,49.5) (25,49.75) (30,47.75)};
  \addplot coordinates{(10,49.14) (15,48.05) (20,48.36) (25,48.44) (30,49.21)};
\legend{5-driver, 10-driver}
\end{axis}
\end{tikzpicture} }
\subfigure[ResLSTMARNet]{
\begin{tikzpicture}[xscale=0.75, yscale=0.6]
  \begin{axis}[ybar,
  symbolic x coords={10,15,20,25,30},xtick=data,
  legend pos=outer north east]
  \addplot coordinates {(10,48) (15,46.25) (20,47) (25,45.35) (30,46.25)};
  \addplot coordinates {(10,48.43) (15,48.28) (20,48.28) (25,48.35) (30,48.28)};
\legend{5-driver, 10-driver}
\end{axis}
\end{tikzpicture} }
\caption{Average identification accuracy (\%) achieved on truck driving data using different $L_f$ values.}
\label{FIG:tfd_Lf}
\end{figure}
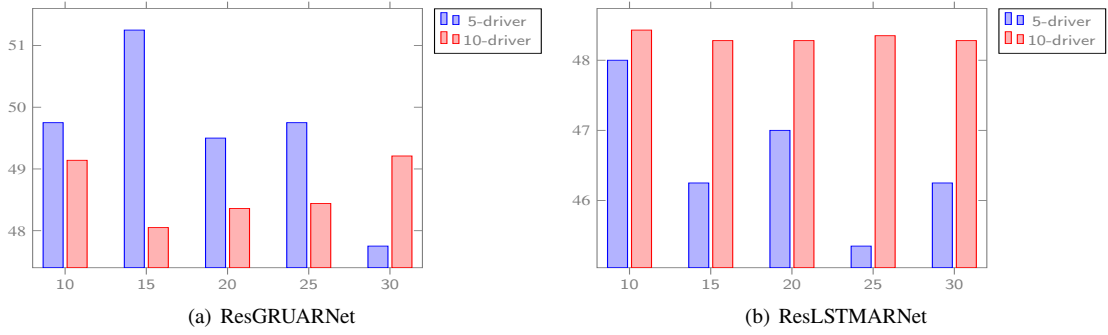

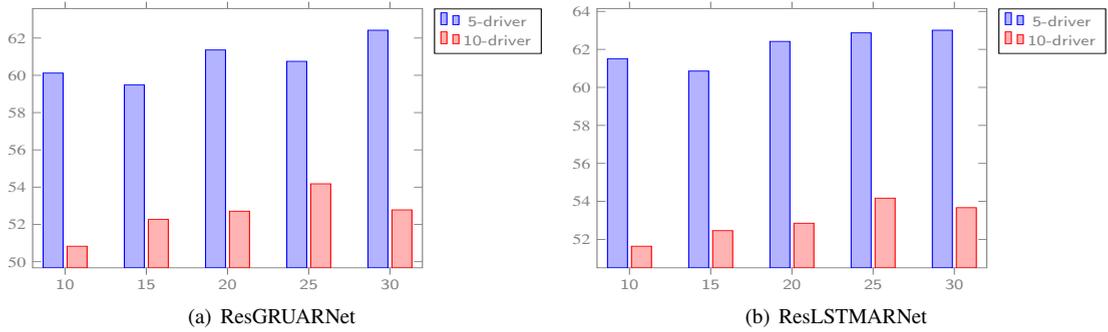
\begin{figure}[htbp]
\centering
\subfigure[ResGRUARNet]{
\begin{tikzpicture}[xscale=0.75, yscale=0.6]
  \begin{axis}[ybar,
  symbolic x coords={10,15,20,25,30},xtick=data,
  legend pos=outer north east]
  \addplot coordinates{(10,60.12) (15,59.49) (20,61.36) (25,60.75) (30,62.42)};
  \addplot coordinates{(10,50.83) (15,52.27) (20,52.71) (25,54.18) (30,52.78)};
\legend{5-driver, 10-driver}
\end{axis}
\end{tikzpicture} }
\subfigure[ResLSTMARNet]{
\begin{tikzpicture}[xscale=0.75, yscale=0.6]
  \begin{axis}[ybar,
  symbolic x coords={10,15,20,25,30},xtick=data,
  legend pos=outer north east]
  \addplot coordinates {(10,61.51) (15,60.86) (20,62.42) (25,62.87) (30,63.01)};
  \addplot coordinates {(10,51.64) (15,52.46) (20,52.85) (25,54.17) (30,53.67)};
\legend{5-driver, 10-driver}
\end{axis}
\end{tikzpicture} }
\caption{Average identification accuracy (\%) achieved on sedan driving data using different $L_f$ values.}
\label{FIG:pcd_Lf}
\end{figure}

Fig.~\ref{FIG:tfd_Lf} shows the results of the truck driving test, where the x coordinates list the options of $L_f$, and the bars in blue and red represent the accuracy of the identification of 5 and 10 drivers. From both bar charts, we note that the three highest bars occur when $L_f$ is 10. However, Fig.~\ref{FIG:pcd_Lf} shows different cases for the sedan driving data. Both models produce the best 10-driver classification accuracies when $L_f$ is 25, but their best performance in the 5-driver group is given when $L_f$ is 30. It is obvious that the segment length poses a modicum of influence on the performance of our models, but the optimal value of $L_f$ depends on specific data. In the following experimental sections, we arbitrarily set $L_f$ to 10 for all models to ensure a fair comparison.

\subsubsection{Soft regularization weight $\lambda$}
Since we introduce a weight term in the proposed regularized loss function, the influence of $\lambda$ on the prediction performance is then studied. In this section, we let $\lambda$ range from 0 to 1 with a step of 0.05 and determine the average prediction accuracy obtained with our two models.

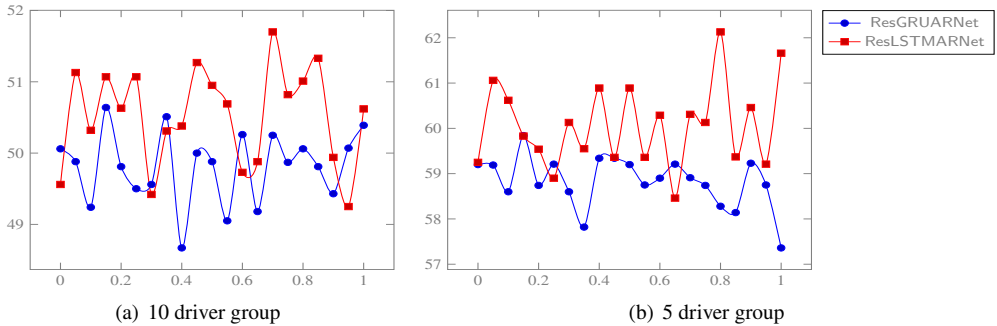
\begin{figure}[htbp]
\centering
\subfigure[10 driver group]{
\label{FIG:lambda_10_driver_pcd}
\begin{tikzpicture}[xscale=0.7, yscale=0.6]
\begin{axis}[legend pos=outer north east]
\addplot+[smooth] 
coordinates{
(0.00,50.06)
(0.05,49.88)
(0.10,49.24)
(0.15,50.64)
(0.20,49.81)
(0.25,49.5)
(0.30,49.56)
(0.35,50.51)
(0.40,48.67)
(0.45,50)
(0.50,49.88)
(0.55,49.05)
(0.60,50.26)
(0.65,49.18)
(0.70,50.25)
(0.75,49.87)
(0.80,50.06)
(0.85,49.81)
(0.90,49.43)
(0.95,50.07)
(1.00,50.39)
};
\addplot+[smooth] 
coordinates{
(0.00,49.56)
(0.05,51.13)
(0.10,50.32)
(0.15,51.07)
(0.20,50.63)
(0.25,51.07)
(0.30,49.42)
(0.35,50.31)
(0.40,50.38)
(0.45,51.27)
(0.50,50.95)
(0.55,50.69)
(0.60,49.73)
(0.65,49.88)
(0.70,51.7)
(0.75,50.82)
(0.80,51.01)
(0.85,51.33)
(0.90,49.94)
(0.95,49.25)
(1.00,50.62)
};
\end{axis}
\end{tikzpicture} }
\subfigure[5 driver group]{
\label{FIG:lambda_5_driver_pcd}
\begin{tikzpicture}[xscale=0.7, yscale=0.6]
\begin{axis}[legend pos=outer north east]
\addplot+[smooth]  
coordinates{
(0.00,59.2)
(0.05,59.19)
(0.10,58.6)
(0.15,59.84)
(0.20,58.74)
(0.25,59.21)
(0.30,58.6)
(0.35,57.82)
(0.40,59.34)
(0.45,59.34)
(0.50,59.2)
(0.55,58.75)
(0.60,58.9)
(0.65,59.21)
(0.70,58.91)
(0.75,58.74)
(0.80,58.28)
(0.85,58.14)
(0.90,59.23)
(0.95,58.75)
(1.00,57.36)
};
\addplot+[smooth]  
coordinates{
(0.00,59.25)
(0.05,61.06)
(0.10,60.62)
(0.15,59.83)
(0.20,59.54)
(0.25,58.9)
(0.30,60.13)
(0.35,59.55)
(0.40,60.89)
(0.45,59.36)
(0.50,60.89)
(0.55,59.36)
(0.60,60.29)
(0.65,58.46)
(0.70,60.31)
(0.75,60.13)
(0.80,62.13)
(0.85,59.37)
(0.90,60.46)
(0.95,59.21)
(1.00,61.66)
};
\addlegendentry{ResGRUARNet}
\addlegendentry{ResLSTMARNet}
\end{axis}
\end{tikzpicture} }
\caption{The average accuracy (\%) of the proposed models on sedan driving data with increasing $\lambda$ values.}
\label{FIG:lambda_pcd}
\end{figure}

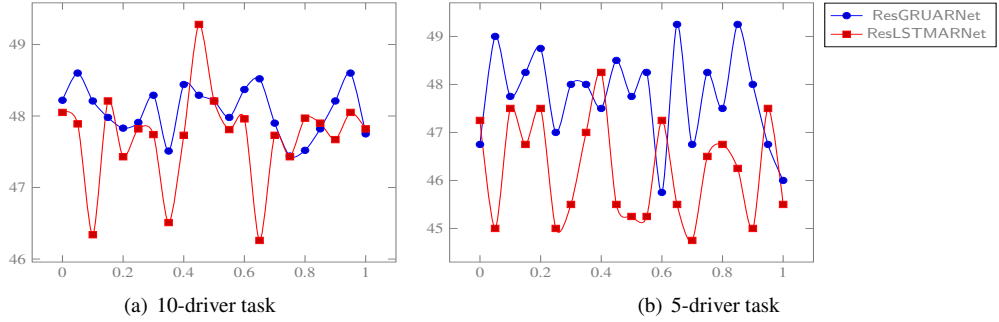
\begin{figure}[htbp]
\centering
\subfigure[10-driver task]{
\label{FIG:lambda_10_driver_tcd}
\begin{tikzpicture}[xscale=0.7, yscale=0.6]
\begin{axis}[legend pos=outer north east]
\addplot+[smooth] 
coordinates{
(0.00,48.22)
(0.05,48.6)
(0.10,48.21)
(0.15,47.98)
(0.20,47.83)
(0.25,47.91)
(0.30,48.29)
(0.35,47.51)
(0.40,48.44)
(0.45,48.29)
(0.50,48.21)
(0.55,47.98)
(0.60,48.37)
(0.65,48.52)
(0.70,47.9)
(0.75,47.44)
(0.80,47.52)
(0.85,47.82)
(0.90,48.21)
(0.95,48.6)
(1.00,47.75)
};
\addplot+[smooth] 
coordinates{
(0.00,48.05)
(0.05,47.89)
(0.10,46.34)
(0.15,48.21)
(0.20,47.43)
(0.25,47.82)
(0.30,47.74)
(0.35,46.51)
(0.40,47.73)
(0.45,49.28)
(0.50,48.21)
(0.55,47.81)
(0.60,47.96)
(0.65,46.26)
(0.70,47.73)
(0.75,47.43)
(0.80,47.97)
(0.85,47.9)
(0.90,47.67)
(0.95,48.05)
(1.00,47.82)
};
\end{axis}
\end{tikzpicture} }
\subfigure[5-driver task]{
\label{FIG:lambda_5_driver_tcd}
\begin{tikzpicture}[xscale=0.7, yscale=0.6]
\begin{axis}[legend pos=outer north east]
\addplot+[smooth]  
coordinates{
(0.00,46.75)
(0.05,49)
(0.10,47.75)
(0.15,48.25)
(0.20,48.75)
(0.25,47)
(0.30,48)
(0.35,48)
(0.40,47.5)
(0.45,48.5)
(0.50,47.75)
(0.55,48.25)
(0.60,45.75)
(0.65,49.25)
(0.70,46.75)
(0.75,48.25)
(0.80,47.5)
(0.85,49.25)
(0.90,48)
(0.95,46.75)
(1.00,46)
};
\addplot+[smooth]  
coordinates{
(0.00,47.25)
(0.05,45)
(0.10,47.5)
(0.15,46.75)
(0.20,47.5)
(0.25,45)
(0.30,45.5)
(0.35,47)
(0.40,48.25)
(0.45,45.5)
(0.50,45.25)
(0.55,45.25)
(0.60,47.25)
(0.65,45.5)
(0.70,44.75)
(0.75,46.5)
(0.80,46.75)
(0.85,46.25)
(0.90,45)
(0.95,47.5)
(1.00,45.5)
};
\addlegendentry{ResGRUARNet}
\addlegendentry{ResLSTMARNet}
\end{axis}
\end{tikzpicture} }
\caption{The average accuracy (\%) of the proposed models on truck driving data with increasing $\lambda$ values.}
\label{FIG:lambda_tcd}
\end{figure}

Fig.~\ref{FIG:lambda_pcd} shows the results for the sedan driving data. The influence of $\lambda$ on the truck driving data is shown in Fig.~\ref{FIG:lambda_tcd}. In these plots, the x-coordinate represents the weight values, and each line corresponds to a type of model. First, regardless of the model used, the influence of weight values on prediction performance is not obvious. Therefore, those lines will neither gradually increase nor decrease with increasing weight.

For 10-driver identification on sedan data, the lines corresponding to ResGRUARNet and ResLSTMARNet in Fig.~\ref{FIG:lambda_10_driver_pcd} reach their peak when $\lambda$ is 0.15 and 0.70, respectively. Meanwhile, for the 5-driver group, the corresponding lines in Fig.~\ref{FIG:lambda_5_driver_pcd} reach their peaks when $\lambda$ is 0.15 and 0.85. We also note that the performance of the ResLSTMARNet model appears to be generally better than that based on GRU. However, the reverse situation is observed in Fig.~\ref{FIG:lambda_tcd} for truck driving data. Specifically, $\lambda$ takes 0.05 and 0.45 to contribute to the highest accuracy in Fig.~\ref{FIG:lambda_10_driver_tcd} for ResGRUARNet and ResLSTMARNet, respectively. For the 5-driver identification task, the two lines in Fig.~\ref{FIG:lambda_5_driver_tcd} reach their peak when $\lambda$ is 0.65 and 0.4.

Observing the influence of $\lambda$ on the performance of the proposed network relying on the LSTM or GRU, we conclude that (1) the regularization weight for the reconstruction loss is necessary since most of the highest accuracy values are obtained when $0< \lambda < 1$. (2) The model performance is sensitive to the choice of this weight term with respect to both the types of RNN and the input patterns. Therefore, this parameter should be properly tuned, and even the extreme case of $\lambda$=0 or 1 could be considered.

\subsection{Ablation study}

\subsubsection{Robustness of pattern fusion}
In this section, we compare different types of local driving patterns to prove the effectiveness of pattern fusion. Given the optimal reconstruction loss weights derived in the last section, we let ResGRUARNet and ResLSTMARNet receive input with MS, ST, and MS+ST patterns and study their resulting identification performance.

The bar charts in Fig.~\ref{FIG:pcd_10diver_patterns} illustrate the average classification accuracies on the sedan driving data. As we can see, there is a significant gap between the bars on MS and ST. The dominant pattern for 5 drivers (Fig.~\ref{FIG:pcd_5diver_pattern_top1}) is MS, but that for 10 drivers (Fig.~\ref{FIG:pcd_10diver_pattern_top1}) is ST. Moreover, the difference occurs even under the same model. However, the two models trained with MS+ST performed well on both tasks, even better than the model trained with a single pattern.

\begin{figure}[htbp]
\centering
\subfigure[5-driver group]{
\label{FIG:tfd_5diver_pattern_top1}
\begin{tikzpicture}[xscale=0.8, yscale=0.8]
  \begin{axis}[ybar,
  symbolic x coords={MS,ST,MS+ST},xtick=data, legend style={at={(0.5,1.0)},anchor=north}]
  \addplot coordinates{(MS,49.25) (ST,43) (MS+ST,50.00) }; 
  \addplot coordinates{(MS,51) (ST,43.75) (MS+ST,48.25) }; 
\legend{GRU, LSTM}
\end{axis}
\end{tikzpicture} }
\subfigure[10-driver group]{
\label{FIG:tfd_10diver_pattern_top1}
\begin{tikzpicture}[xscale=0.8, yscale=0.8]
  \begin{axis}[ybar,
  symbolic x coords={MS,ST,MS+ST},xtick=data, legend style={at={(0.5,1.0)},anchor=north}]
  \addplot coordinates{(MS,48.6) (ST,33.7) (MS+ST,48.6) }; 
  \addplot coordinates{(MS,48.7) (ST,34.1) (MS+ST,49.28) }; 
\legend{GRU, LSTM}
\end{axis}
\end{tikzpicture} }
\caption{Average identification accuracy (\%) achieved on truck driving data using different driving patterns.}
\label{FIG:tfd_10diver_patterns}
\end{figure}
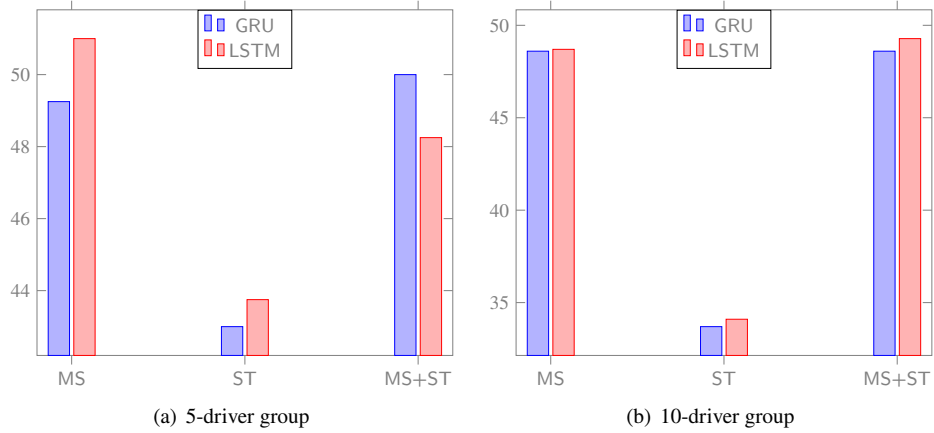

Fig.~\ref{FIG:tfd_10diver_patterns} shows the comparison results on truck driving data. For both tasks, the MS pattern, rather than ST, occupies the dominant place to help both models exhibit better results. Similar or even better results can also be observed when using MS+ST in both subfigures.

The two figures indicate that the driving behavior between trucks and sedans is different, as reflected in the MS and ST patterns. Furthermore, the use of MS or ST alone likely results in poor identification performance when the learned model is applied to data sets from different types of vehicles. To address the third drawback, a more stable and powerful local representation ability can be achieved by combining ST and MS together.

The study in this section shows that (1) the fusion of ST and MS patterns effectively captures the segment-level driving patterns of different drivers. Pattern fusion is more robust to the two data sets. (2) Both LSTM and GRU are chosen here to show similar trends, which indicates that the effect of the local pattern is independent of a specific RNN layer.

\begin{figure}[htbp]
\centering
\subfigure[5-driver group]{             
\label{FIG:pcd_5diver_pattern_top1}
\begin{tikzpicture}[xscale=0.8, yscale=0.8]
\begin{axis}[ybar,
  symbolic x coords={MS,ST,MS+ST},xtick=data, legend style={at={(0.5,1.0)},anchor=north}] %
  \addplot coordinates{(MS,61.5) (ST,54) (MS+ST,61.5) };  
  \addplot coordinates{(MS,61.5) (ST,52) (MS+ST,62.13) }; 
\legend{ResGRUARNet, ResLSTMARNet}
\end{axis}
\end{tikzpicture} }
\subfigure[10-driver group]{        
\label{FIG:pcd_10diver_pattern_top1}
\begin{tikzpicture}[xscale=0.8, yscale=0.8]
  \begin{axis}[ybar,
  symbolic x coords={MS,ST,MS+ST},xtick=data, legend pos=north west]
  \addplot coordinates{(MS,41) (ST,52) (MS+ST,51) };  
  \addplot coordinates{(MS,41) (ST,51.5) (MS+ST,51.7) };  
\legend{ResGRUARNet, ResLSTMARNet}
\end{axis}
\end{tikzpicture} }
\caption{Average identification accuracy (\%) achieved on sedan driving data using different driving patterns.}
\label{FIG:pcd_10diver_patterns}
\end{figure}
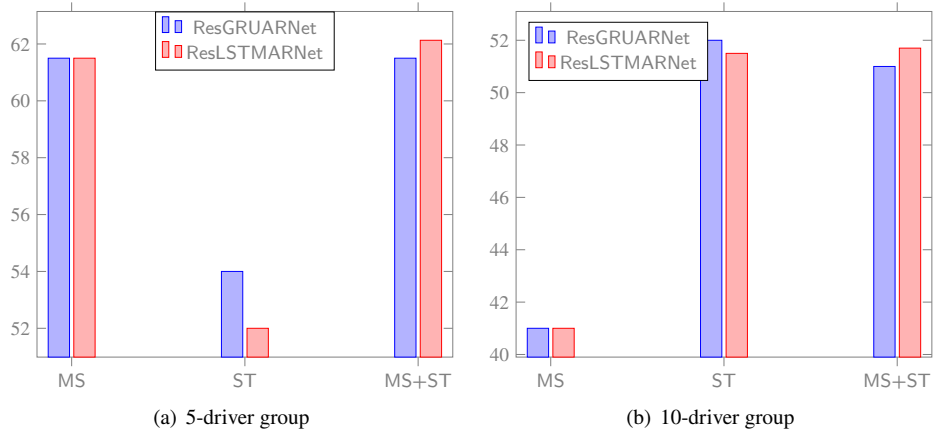

\subsubsection{Effectiveness of the proposed network}
The highlights of our proposed network structure are residual connections and the design of the autoencoder. To study the effectiveness of this design, we first remove the decoder component to obtain the residual RNN classifier. Then, the skip connection and the first RNN block are removed from the residual RNN classifier to obtain a basic RNN classifier. By replacing the RNN with the LSTM and GRU, we consider the basic LSTM and GRU models and residual LSTM and GRU models (denoted ResLSTM and ResGRU, respectively) for comparison.

First, we apply the decomposed models and our proposed models to the sedan driving data. The average accuracy values (\%) for the identification tasks with 5 and 10 drivers are listed in Table~\ref{tab:sedan_net_ablation}. When comparing the accuracy values from a vertical view, it can be observed that the identification accuracy on both tasks gradually increases, regardless of the RNN type.

\begin{table}
\caption{Comparison of the identification accuracy (\%) to the decomposed models tested on sedan driving data.}\label{tab:sedan_net_ablation}
\begin{tabular}{cccccc}
\hline
\textbf{GRU-based models}    & \textbf{5-driver}    & \textbf{10-driver} &\textbf{LSTM-based models}    & \textbf{5-driver}    & \textbf{10-driver} \\ \hline
GRU         & 57.69    & 43.60 & LSTM        & 60.57    & 47.39 \\ 
ResGRU      & 59.81    & 49.18 & ResLSTM     & 61.20     & 49.36 \\ 
ResGRUARNet & 59.84    & 50.64 & ResLSTMARNet & 62.13   & 51.70 \\ \hline
\end{tabular}
\end{table}

We then apply the decomposed models and our proposed models to the truck driving data and obtain their average accuracy. The results are listed in Table~\ref{tab:truck_net_ablation}. Despite the exception that the accuracy of the residual LSTM model is slightly lower than that of the basic LSTM model, we again note the same effect: the residual connection improves the performance of the basic RNN classifiers, and the performance of residual RNN models is further improved by the decoder component.

Both tables show that the design of our network structure is capable of improving the accuracy of identification. In other words, introducing the residual autoencoder (or the additional weighted loss) to the traditional RNN classifier actually helps improve driver identification performance. In the next section, we further demonstrate the superiority of our proposed framework over other related models.

\begin{table}
\caption{Comparison of the identification accuracy (\%) to the decomposed models tested on truck driving data.}\label{tab:truck_net_ablation}
\begin{tabular}{cccccc} 
\hline
\textbf{GRU-based models}    & \textbf{5-driver}    & \textbf{10-driver} &\textbf{LSTM-based models}    & \textbf{5-driver}    & \textbf{10-driver} \\ \hline
GRU         & 46.00    & 45.35 & LSTM        & 46.00    & 46.42 \\
ResGRU      & 47.50    & 48.60 & ResLSTM     & 45.25    & 47.82 \\
ResGRUARNet & 49.25    & 48.60  & ResLSTMARNet & 48.25   & 49.28  \\ \hline
\end{tabular}
\end{table}

On the basis of the ablation study results, we experimentally prove that the fusion pattern (MS+ST) is more robust for characterizing the driving behavior of drivers of both sedans and trucks. With such a fused pattern, the proposed ResRNNARNet proves the necessity of combining residual connections and autoencoder regularization tricks to further learn strong driving-style representations and exhibits better identification performance.

\subsection{Performance comparison}
\label{sec:comp}
In this section, we train different candidate models and compare their driver classification performance with that of our proposals. The codes for all the models are available in the Git repository in the footnote\footnote{https://gitee.com/cs\_icv/driver\_identification} for reproducibility. Note that the reproduced results may exhibit slight fluctuations due to randomness, but the superiority of our solution can still be achieved. Since some drivers may have similar driving styles, our model learns only from GPS data and is incapable of in-depth differentiation. Therefore, the top-3 accuracy is adopted for basic recognition and is used in some references as well. Furthermore, to demonstrate the reliability of the model performance, we also collected test results from the previously mentioned 5 repetitions of 5-fold cross-validation and calculated the 95\% confidence interval (CI). The comparison methods are as follows.

\textbf{CNN Model} (adapted). This is a CNN classifier proposed by \cite{Dong2016,hu2020driver}. Using statistical features as input, this model employs three convolutional layers on top of the input, where each is followed by a max-pooling layer. The original code is available at the link in the footnote~\footnote{https://github.com/sobhan-moosavi/CharacterizingDrivingStylesWithDeepLearning}. To ensure a fair comparison, we adapted the code to allow this model to be fed with fusion-pattern-based input as well.

\textbf{StackedIRNN Model}. This is an RNN model proposed by ~\cite{Dong2016}. This model uses two LSTM blocks with the same dimensions as our proposed model. The original input is an MS matrix, which will be replaced by an MS+ST matrix. The original code is still available through the link in the footnote~\footnote{https://github.com/sobhan-moosavi/CharacterizingDrivingStylesWithDeepLearning}.

\textbf{ARNet Model}. This model, proposed by \cite{Dong2017}, is an autoencoder-based NN model used to build a latent representation for an input subtrajectory and to properly classify trajectories, where the class labels are driver identities. It also introduced an autoencoder to reconstruct the latent vector for hard regularization, that is, $\lambda$ was arbitrarily set to 1. The original codes of this method are available at the link in the footnote~\footnote{\ https://github.com/sobhan-moosavi/ARNet}.

\textbf{Attention-GRU and attention-LSTM}. The attention mechanism is adopted by \cite{zhang2019deep} for the analysis of driver behavior. In their study, stacked GRU/LSTM with two attention-based layers was proposed. We reproduced and adjusted the two models to accommodate our settings.

\textbf{FCN-LSTM Model} (adapted). The fully convolutional LSTM (FCN-LSTM) model was first introduced by \cite{karim2019multivariate} for the task of time series classification and then used for the task of identifying the driver by \cite{8746156}. There are two different blocks in the architecture of FCN-LSTM, including convolutional and LSTM blocks. We make small adaptations based on the original open source code~\footnote{https://github.com/titu1994/LSTM-FCN}. The model is also adapted to receive the same input.

\textbf{D-CRNN}. The method was recently proposed by \cite{moosavi2021driving}. The code is also open source. In the original study, they only used statistical features to train the model, and we adjusted this model to accept the features of the fused pattern.

\textbf{Siamese-LSTM}(adapted). Siamese-LSTM was also recently considered by~\cite{8813795,9716898} to learn driving style embeddings. We first train such a Siamese-LSTM model, and then the driver embeddings are fed to a fully connected layer and softmax for classification. The key margin factor in contrast loss is chosen from 0 to 1 with an increasing step of 0.05.

Table~\ref{tab:sedan_driver_compare} summarizes the identification metrics of all models obtained on sedan driving data. The model that performed best of the first eight compared models is the StackedIRNN. Additionally, other comparison models have poor metrics on the 10-driver classification task.
On the other hand, our two models outperformed them with relatively low confidence intervals. In particular, the ResLSTMARNet model produces the highest top-1 and top-3 accuracies for both tasks, and the advantages are obvious.
 
\begin{table}[]
\caption{Comparison of the identification metrics obtained on sedan driving data.}\label{tab:sedan_driver_compare}
\begin{tabular}{lllll}
\hline
\multicolumn{1}{c}{}        & \multicolumn{2}{c}{5-driver}           & \multicolumn{2}{c}{10-driver}              \\
\multicolumn{1}{l}{Models} & \multicolumn{1}{l}{top-1$\pm$CI(\%)} & \multicolumn{1}{c}{top-3$\pm$CI(\%)} & \multicolumn{1}{c}{top-1$\pm$CI(\%)} & top-3$\pm$CI(\%) \\ \hline
CNN                 & 57.51$\pm$2.43    & 80.55$\pm$3.91    & 29.37$\pm$1.63    & 55.64$\pm$1.61  \\ 
StackedIRNN         & 58.15$\pm$2.79    & 86.32$\pm$5.42    & 49.24$\pm$1.86    & 68.25$\pm$3.08  \\ 
ARNet               & 49.60$\pm$2.04    & 80.24$\pm$3.86    & 29.12$\pm$1.61    & 46.07$\pm$1.83  \\ 
LSTM-FCN            & 49.19$\pm$2.84    & 81.31$\pm$4.89    & 42.73$\pm$4.16    & 68.24$\pm$4.64  \\ 
Attention-GRU       & 50.79$\pm$2.76    & 81.90$\pm$4.52    & 39.62$\pm$2.92    & 57.91$\pm$1.90  \\ 
Attention-LSTM      & 49.60$\pm$2.04    & 79.92$\pm$4.06    & 29.12$\pm$1.61    & 47.98$\pm$2.36  \\ 
Siamese-LSTM        & 31.86$\pm$6.53    & 73.40$\pm$5.04    & 16.94$\pm$4.69    & 37.20$\pm$5.79  \\ 
DCRNN               & 41.99$\pm$5.41    & 77.05$\pm$5.41    & 25.75$\pm$3.71    & 48.05$\pm$2.62  \\ 
ResGRUARNet         & \textbf{59.84$\pm$2.08}    & \textbf{86.99$\pm$5.24}    & \textbf{50.64$\pm$2.40}    & \textbf{74.88$\pm$2.58}  \\
ResLSTMARNet        & \textbf{62.13$\pm$2.36}    & \textbf{87.89$\pm$5.44}    & \textbf{51.70$\pm$2.22}    & \textbf{76.26$\pm$2.70}   \\\hline
\end{tabular}
\end{table}

Next, we investigate the identification metrics obtained from the truck driving data in Table~\ref{tab:truck_driver_compare}.
For the 5-driver identification task, in terms of top-1 accuracy, the attention-GRU model turns out to be the best. This is followed by the Siamese-LSTM model. However, the corresponding high confidence intervals challenge the reliability of the two models. The two models are also the best in terms of top-3 accuracy. Although our models are not the best at this task, they still achieve good performance. Moreover, powerful robustness and generalizability can be observed considering their performance on both tasks. For the more challenging 10-driver identification task, we once again note the poor performances of the compared methods except for the StackedIRNN. Attention-GRU and Siamese-LSTM lose their superiority. On the other hand, this phenomenon highlights the advantages of our models. Our models produce the best top-1 and top-3 accuracies, and the confidence intervals reflect that their reliability is strong.

A comparison of Table~\ref{tab:sedan_driver_compare} and Table~\ref{tab:truck_driver_compare} indicates that existing models have their advantages; however, their capabilities cannot be transferred to different driving data sets. In contrast, our approach achieves state-of-the-art performance in most cases. It can be concluded that our proposed residual recurrent autoencoder framework, using fused local patterns, outperforms existing deep learning methods when using only GPS-derived data. However, we believe that determining the optimal RNN structure still depends on the specific task.

\begin{table}[]
\caption{Comparison of the identification metrics obtained on truck driving data.}\label{tab:truck_driver_compare}
\begin{tabular}{lllll}
\hline
\multicolumn{1}{c}{}        & \multicolumn{2}{c}{5-driver}      & \multicolumn{2}{c}{10-driver}              \\ 
\multicolumn{1}{l}{Models} & \multicolumn{1}{l}{top-1$\pm$CI(\%)} & \multicolumn{1}{c}{top-3$\pm$CI(\%)} & \multicolumn{1}{c}{top-1$\pm$CI(\%)} & top-3$\pm$CI(\%) \\ \hline
CNN                 & 49.25$\pm$4.61    & 83.00$\pm$1.90    & 38.89$\pm$2.01    & 60.14$\pm$2.34  \\ 
StackedIRNN         & 43.75$\pm$3.73    & 78.50$\pm$2.60    & 46.42$\pm$1.78    & 65.48$\pm$2.32  \\ 
ARNet               & 49.50$\pm$6.89    & 83.50$\pm$1.95    & 33.72$\pm$2.06    & 57.80$\pm$1.83  \\ 
LSTM-FCN            & 30.00$\pm$8.37    & 68.50$\pm$8.18    & 40.08$\pm$2.65    & 65.16$\pm$2.40  \\ 
Attention-GRU       & \textbf{52.50$\pm$5.92}    & \textbf{83.50$\pm$1.95}    & 35.80$\pm$2.54    & 62.92$\pm$2.86  \\ 
Attention-LSTM      & 43.75$\pm$3.73    & 81.50$\pm$2.64    & 33.72$\pm$2.06    & 58.58$\pm$2.12  \\ 
Siamese-LSTM        & \textbf{51.75$\pm$5.39}    & \textbf{84.75$\pm$1.97}    & 17.53$\pm$5.01    & 39.73$\pm$4.81  \\ 
DCRNN               & 38.25$\pm$8.52    & 78.75$\pm$3.73    & 28.37$\pm$3.80    & 52.05$\pm$3.69  \\ 
ResGRUARNet         & 50.00$\pm$4.16    & 79.75$\pm$2.80    & \textbf{48.60$\pm$1.23}    & \textbf{69.12$\pm$1.93}  \\
ResLSTMARNet        & 48.25$\pm$4.78    & 81.75$\pm$2.76    & \textbf{49.28$\pm$2.03}    & \textbf{68.64$\pm$2.29}   \\\hline
\end{tabular}
\end{table}

\section{Conclusion}
\label{sec_6}
In this study, we integrated the MS and ST domains to characterize the kinematic segments of vehicular movements with the aim of identifying the driver. This fusion yielded more information that could help characterize the short-term operations of the driver of a vehicle. The fusion of patterns from the two domains better captured local driving patterns than either domain alone.

We then used a residual recurrent autoencoder-regularized network to learn the driving style over a fixed-length subtrajectory. Subsequently, a unified learning strategy was proposed that regularizes the prior classification network with an autoencoder to improve the generalization of the representation of the learned driving style. The 5- and 10-driver classification tasks conducted on two data sets showed that our proposed types of input fed to our networks outperformed the benchmark methods. In view of the above results, this calls for the necessity of combining both MS and ST patterns to better capture segment-level driving behaviors.

Finally, our solution relies on easy-to-calculate and explainable statistical and state transitional features for robustness. The proposed networks are designed to learn better driving style representations to improve identification performance. However, extra effort is needed to fine-tune $\lambda$. In ongoing work, we will try to consider ways to improve the classification accuracy of our models for many drivers. Identifying solutions such as considering more data sources and advanced networks for these problems is left for future work.


\printcredits

\bibliographystyle{cas-model2-names}
\bibliography{cas-refs}




\end{document}